\documentstyle[12pt]{article}
\begin{document}
\title{The $(g-2)_{\mu}$ anomaly, Higgs bosons and heavy neutrinos}
\author{G.G.Boyarkina, O.M.Boyarkin\thanks{E-mail: oboyarkin@btut.by}\\
Byelorussian State Pedagogical University\\
Soviet Street 18, Minsk, 220050, Belarus}
\date {*}
\maketitle
\begin {abstract}
Within the model based on the $SU (2) _L\times SU (2) _R\times U (1) _ {B-L} $
gauge group and having the bidoublet and two triplets of the Higgs fields
(left-right model) the Higgs sector impact on the value of the muon
anomalous magnetic moment (AMM) is considered.
The contributions coming from the doubly charged Higgs bosons, the singly
charged Higgs bosons and the lightest neutral Higgs boson are taken into
account. The obtained value of the muon €ŒŒ is the function of
the Higgs boson masses and the Higgs boson couplings constants (CC's).
We express the most of part of the CC's as the functions of
the heavy neutrino sector parameters. We show that at the particular parameters
values the model under study could explain the BNL'00 result. \\
PACS number(s): +12.15.Cc, 12.15.Ff,-13.40.Em.
\end {abstract}

\section {Introduction}
Measurements of the spin magnetic dipole moment of particles have
a rich history as harbingers of impressive progress in the quantum
theory. Thus, registration of the anomalous values of the nucleons
magnetic moments was powerful argument for the benefit of the $\pi
$-mesonic theory of the  nuclear forces formulated by Yukawa.
Determination of the anomalous magnetic moment (AMM) of the
electron has played an important role in development of modern
quantum electrodynamics and renormalization theory. It appeared to
be reasonable that the ongoing muon $(g-2)_{\mu}$ measurement E821
at Brookhaven National Laboratory (BNL) on Alternating Gradient
Synchrotron would be a sensitive test for the results of the
standard model (SM) electroweak corrections. Because this
experiment is the culmination of a series of measurements of
ever-increasing accuracy over the past several years then its
results has triggered the interest in theoretical calculations of
the muon AMM. The first BNL result based on the data taken through
1997 was [1] $$a_{\mu}^{exp}=(116\ 592\
500\pm1500)\times10^{-11}\mu_0\qquad(\mbox{BNL'97}), \eqno (1) $$
where $\mu_0$ is the muon magnetic moment predicted by the Dirac's
theory. The 1998 and 1999 runs had the much higher statistics and
gave the results with the increased precession
$$a_{\mu}^{exp}=(116\ 591\ 910\pm590)\times10^{-11}\mu_0
\qquad(\mbox{BNL'98} \ [2]).\eqno (2)$$ $$a_{\mu}^{exp}=(116\ 592\
020\pm160)\times10^{-11}\mu_0 \qquad(\mbox{BNL'99} \ [3]).\eqno
(3)$$ The BNL'98 and BNL'99 results averaged with older
measurements made at CERN [4] brought to the following value of
the muon AMM $$a_{\mu}^{exp}=(116\ 592\
023\pm151)\times10^{-11}\mu_0, \eqno(4)$$

In the SM the expression for the muon AMM can be presented as a sum
$$a_{\mu}^{SM}=a_{\mu}^{QED}+a_{\mu}^{EW}+
a_{\mu}^{had},\eqno(5)$$
in which $a_{\mu}^{QED}=11\ 658\ 470.57(0.29)\times10^{-10}\mu_0$ (see [5]
and references therein) and
$a_{\mu}^{EW}=15.2(0.4)\times10^{-10}$ (see [6] and references therein).

The term $a_{\mu}^{had}$  arises from virtual hadronic contributions to the
photon propagator in $4^{th}$ $a_{\mu}^{had}(VP1)$ and $6^{th}$ order, where
the latter includes hadronic vacuum polarization $a_{\mu}^{had}(VP2)$ and
light-by-light scattering $a_{\mu}^{had}(LbyL)$.
The dominant contribution to $a_{\mu}^{had}$ as well as one of the
largest ambiguities in its value come from the $a_{\mu}^{had}(VP1)$.
The $a_{\mu}^{had}(VP1)$ has been derived in Ref. [7] from the $e^+e^-$
hadronic cross section and the hadronic $\tau$ decay data
$$a_{\mu}^{had}(VP1)=6\ 924(62)\times10^{-11}.\eqno(6)$$
Evolution of 3-loop hadron vacuum polarization contribution
$a_{\mu}^{had}(VP2)$ has given the result [8]
$$a_{\mu}^{had}(VP2)=-100(6)\times10^{-11}\mu_0.\eqno(7)$$

It is important to keep in mind that all the estimations of the
$LbyL$ scattering contribution $a_{\mu}^{had}(LbyL)$ made so far
are the model dependent. The calculations are based on the chiral
perturbation or extended Nambu-Jona-Lasinio model. Also the vector
meson dominance is assumed and the phenomenological
parametrization of the pion form factor $\pi\gamma^*\gamma^*$ is
introduced in order to regularize the divergence. The previous
average value for $a_{\mu}^{had}(LbL)$ is given by [9,10]
$$a_{\mu}^{had}(LbyL)=-85(25)\times10^{-11}\mu_0.\eqno(8)$$ With
this value of the $LbyL$ hadronic correction the total SM
prediction of $a_{\mu}^{SM}$ was $$a_{\mu}^{SM}=116\ 591\ 597
(67)\times10^{-11}\mu_0.\eqno(9)$$ Comparing Eq. (9) with the
experimental average in Eq. (4) one could find $$\delta
a_{\mu}\equiv a_{\mu}^{exp}-a_{\mu}^{SM}=426(165)\times10^{-11}
\mu_0.\eqno(10)$$ Eq. (10) means that there is the 2.6 $\sigma$
deviation between experiment and the SM prediction.

Recently the theoretical prediction for the $a_{\mu}^{had}(LbyL)$
has undergone a significant revision because of the change in
sign. The re-evaluations have given the following values for
$a_{\mu}^{had}(LbyL)$
$$a_{\mu}^{had}(LbyL)=\left\{\begin{array}{lll}
83(12)\times10^{-11}\mu_0 \qquad [11],\\ 89(15)\times10^{-11}\mu_0
\qquad [12],\\ 83(32)\times10^{-11}\mu_0\qquad
[13].\end{array}\right.\eqno(11)$$ Taking the average of these new
results one finds $$a_{\mu}^{SM}=116\ 591\ 770
(70)\times10^{-11}\mu_0.\eqno(12)$$ Using Eqs. (4) and (12) one
could obtain $$\delta
a_{\mu}=260(160)\times10^{-11}\mu_0.\eqno(13)$$ Thus the deviation
value has droped from 2.6 $\sigma$ up to 1.6 $\sigma$.

On July 30, 2002 Muon g-2 Collaboration announced the new result based on the
$\mu^+$ data collected in the year 2000 [14]
$$a_{\mu}^{exp}=116\ 592\ 040(70)(50)\times10^{-11}\mu_0.\qquad (\mbox{BNL'00})
\eqno(14)$$
An uncertainty of BNL'00 is almost two times smaller than in BNL'99
and only two times larger than the final aim of the E821 experiment.
With this new result the present world average experimental value is
$$a_{\mu}^{exp}=116\ 592\ 030(80)\times10^{-11}\mu_0.\eqno(15)$$

The improved calculations of the $a_{\mu}^{had}(VP1)$  have been presented
recently [15,16]. These are data-driven analysis using the most recent data
from the $e^+e^-$ hadronic cross section observed at CMD-2, BES, SND [17].
Their precision $\sim58\times10^{-11}$ are now even smaller than those
in Eq. (6). Further on
for the $a_{\mu}^{had}(VP1)$ we shall use the result of Ref. [15] where
the experimental input is based only on the $e^+e^-$ data
$$a_{\mu}^{had}(VP1)=6\ 889(58)\times10^{-11}\mu_0.\eqno(16)$$
For the estimation of the $a_{\mu}^{had}(LbyL)$ we invoke the new result
obtained in [18]
$$a_{\mu}^{had}(LbyL)=80(40)\times10^{-11}\mu_0.\eqno(17)$$
Then with help of Eqs. (7), (16) and (17) the full hadronic contributions
is given by
$$a_{\mu}^{had}=6\ 869(71)\times10^{-11}\mu_0.\eqno(18)$$
This leads us to the SM prediction
$$a_{\mu}^{SM}=116\ 591\ 726.7(70.9)\times10^{-11}\mu_0.\eqno(19)$$
So, at present the deviation between experimental data and the SM
prediction reached the value
$$\delta a_{\mu}=303.3(106.9)\times10^{-11}\mu_0,\eqno(20)$$
that is, the deviation is roughly about 3$\sigma$.

Since the E821 data have been thoroughly collected and studied
over many years, it is most unlikely that this discrepancy could
be also explained as a mere statistical fluctuation, as several
earlier deviations from the SM turned out to be. Attention is
drawn to the fact of the extremely small variation of the muon AMM
central value in all the BNL results presented up to now. This
circumstance could be the weighty argument in favour of a
trustworthiness of the E821 experiment. While it is often argued
that the SM should be augmented by New Physics at higher energy
scales because of some unanswered fundamental questions, the
$(g-2)_{\mu}$ anomaly with such phenomena as the neutrino
oscillations [19], 3$\sigma$ departure of $\sin^2\theta_W $ from
the SM predictions measured in the deep inelastic neutrino-nucleon
scattering [20], and the observation of the neutrinoless double
beta decay [21] may serve as the New Physics signal already at the
weak scale. If the deviation of Eq. (20) can be attributed to
effects of the physics beyond the SM, then at 95\% $CL$, $\delta
a_{\mu}/\mu_0$ must lie in the range
$$93.8\times10^{-11}\leq{\delta
a_{\mu}\over\mu_0}\leq512.8\times10^{-11}. \eqno(21)$$ This
contribution is positive, and has the same order as the
electroweak corrections to $a_{\mu}$, namely $\sim
G_Fm_{\mu}^2/(4\pi^2\sqrt{2}).$

Suggestions already made
in literature for explaining $\delta a_{\mu}/\mu_0$ include
supersymmetry [22], additional gauge bosons
[23], anomalous gauge boson couplings [24], leptoquarks [25], extra dimensions
[26], muon substructure [27], exotic flavour-changing interactions [28],
exotic vectorlike fermions [29], possible nonpertubative effects at the
1 TeV order [30] and so on.

Amongst explanations of E821-experiment at Brookhaven AGL the approach based
on the possibility of the violation of $CPT$ and Lorentz invariance
(see, for example [31]) should be particularly noted. In E821-experiment
$a_{\mu}$ is determined by measuring the difference $\omega_a$
between the spin precession angular frequency $\omega_s$ and the cyclotron
angular
frequency $\omega_c$ of highly polarized muons in a storage ring with a
uniform magnetic field.
The field strength is determined from the nuclear magnetic resonance (NMR)
frequency of
protons in water, calibrated relative to the free proton NMR frequency
$\omega_p$. The quantity $a_{\mu}$ is determined via
$$a_{\mu}={\omega_a/\omega_p\over\mu_{\mu}/\mu_p-\omega_a/\omega_p},\eqno(22)$$
where the ratio $\mu_{\mu}/\mu_p$ is taken from the measurement by
W.Liu {\it {et al.}} [32]. According to Ref. [31] $CPT$/Lorentz violating
terms in the Lagrangian induce a shift $\delta\omega_a$ to frequency
$\omega_a$. This shift is predicted to be different for positive
and negative muons and to oscillate with the Earth's sidereal frequency.
The level of $CPT$/Lorentz violating effects is characterized by the
dimensionless quantity $r=\omega_a/m_{\mu}$ which interpret $\delta\omega_a$
as an muon energy shift in respect to the rest energy $m_{\mu}$. Already with
the 1999 data set $r$ could be probed down to the level of $0.19\times10^{-22}$.
Nowadays the common believe is that $CPT $ and Lorentz invariance are the
immovable laws of Nature and, therefore, detecting of their violation must
lead to the radical alterations of the contemporary quantum field theory.

Some of explanations of E821-experiment turn out to be excluded
by the current experimental data. To cite some examples. The possibility of
muon substructure can be immediately ruled out since the necessary compositeness
scale of muon should already have been seen in processes involving
high energetic muons at LEP, HERA, and the Tevatron.

For anomalous $W$-boson dipole magnetic moment
$$\mu_W={e\over2m_W}(1+\kappa_{\gamma})$$
the additional one loop contribution to $a_{\mu}$ is given by the expression
$$a_{\mu}(\kappa_{\gamma})\approx{G_Fm_{\mu}^2\over4\sqrt{2}\pi^2}
\ln\left({\Lambda^2\over m_W^2}\right)(\kappa_{\gamma}-1),$$
where $\Lambda$ is the high momentum cutoff required to give a finite result.
For $\Lambda\approx1$ TeV, in order to obtain the accord between theory and
observation one should demand
$$\delta\kappa_{\gamma}\equiv\kappa_{\gamma}-1\approx0.4.$$
However such a big values of $\delta\kappa_{\gamma}$ is already eliminated by
$e^+e^-\rightarrow W^+W^-$ data at LEP II which gives [33]
$$\delta\kappa_{\gamma}=0.08\pm0.17.$$
In this manner, at the moment the $(g-2)_{\mu}$ anomaly plays
the role of an Occam's razor for the existing SM extensions.

The purpose of this work is to investigate the $(g-2)_{\mu}$ anomaly within
the left right model (LRM) based on the gauge group $SU(2)_R\times
SU(2)_L\times U(1)_{B-L}$. One-loop contributions to $a_{\mu}$ from extra gauge
bosons have been calculated in [34]. However, the contribution coming
from the $Z_2$ gauge boson is negative while in order to accommodate the
discrepancy in Eq. (20) the mass value of the $W_2$ gauge boson should lies
around 100 GeV, which is clearly ruled out by direct searches and precission
measurements [33].
In the LRM the Higgs bosons may apply to the role of the following candidates
which may give significant contributions to the muon AMM.

In the SM, the Higgs boson contribution to $a_{\mu}$ is negligible
because $\overline{\mu}\mu h$ coupling is extremely small, namely
$\sim m_{\mu}/v$, where $v$ is a vacuum expectation value being
equal to $246$ GeV. In the LRM the Higgs sector is much richer
than in the SM. It includes four doubly charged scalars
$\Delta^{(\pm\pm)}_{1,2}$, four singly charged scalars $h^{(\pm)}$
and $\tilde{\delta}^{(\pm)}$, four neutral scalars $S_i$
($i=1,2,3,4 $) and two neutral pseudoscalars $P_{1,2} $. The
current experimental data allow some of these Higgs bosons to have
masses around the electrowek scale and couplings of at least
electroweak strength. It is well to bear in mind that amongst the
extensions of the SM the LRM is of special interest because its
Higgs sector contain the elements belonging to other most popular
nowaday models. The presence of the bidoublet in the LRM  causes
the existence of the same physical Higgs bosons as in the two
Higgs doublet modification of the SM (THDM) [35] and in the MSSM
[36]. Owing to the availability of the triplets the LRM has the
Higgs bosons which are present in the model based on the
$SU(3)_L\times U(1)_N$ gauge group [37].

One more a fascinating property of the LRM resides in the fact that the LRM
belongs among the models in which
the Higgs bosons coupling constants (CC's) determining the interaction of the
Higgs bosons both with leptons and with
gauge bosons are connected to the neutrino oscillation parameters (NOP's).
Therefore, in such models the obtained bounds on the Higgs sector parameters
could be extended to the bounds on the NOP's.

The paper is organized as follows. In the next Sect. the one-loop electroweak
corrections to the muon AMM caused by the LRM Higgs bosons are calculated.
There we establish the connection between the CC's and the NOP's. Then
comparing the theoretical and
the experimental values of $a_{\mu}$ we find the bounds on the
Higgs sector parameters which provide in its turn information on the heavy
neutrino masses and the mixing angles. Sec.4 is devoted to analysis of the
results obtained.

\section{Higgs bosons corrections to $a_{\mu}$}

In the LRM the choice of the Yukawa potential has the influence upon the form
of the Lagrangian describing the Higgs boson interactions both with
fermions and gauge bosons. The most general Yukawa potential
${\cal{L}}^g_Y$ has been proposed in [38]. In spite of the fact that
${\cal{L}}^g_Y$ has the very complicated form the diagonalization of the
charged Higgs bosons mass matrix presents no special problems.
However, for the neutral Higgs bosons mass matrix $M_n$ this procedure
could be only realized when some simplifications in ${\cal{L}}^g_Y$ have been
done [39]. For example, the matrix $M_n$ could be diagonalized
at the following conditions (we use the same notation as in Ref.[38])
$$\alpha_1={2\alpha_2k_2\over k_1}, \qquad
\alpha_3={2\alpha_2k_-^2\over k_1k_2}, \qquad
\beta_1={2\beta_3k_2\over k_1},\eqno(23)$$
where $\alpha_{1,2,3}$ and $\beta_{1,2}$ are the constants entering the
Yukawa potential, $k_1$ ¨ $k_2$ are the vacuum expectation values (VEV's) of
the neutral components of the Higgs bidoublet and $k_{\pm}=
\sqrt{k_1^2\pm k_2^2}$ ($k_+=174$ GeV).

Of all the Higgs bosons,
$\Delta^{(\pm\pm)}_{1,2}$-, $h^{(\pm)}$ and $\tilde{\delta}^{(\pm)}$- and
$S_1$-bosons have been of our main interest here because the current data
allow their masses to lie on the electroweak scale (recall that the
$S_1$-boson is the analog of the SM Higgs boson).
Assuming the conditions (23) to be fulfilled one obtain the squared masses of
these particles
$$m_h^2=\alpha(v_R^2+k_0^2)+{\beta_1^2k_+^4k_0^2\over{k_-^4(\alpha +\rho_1-
\rho_3/2)}},\eqno(24)$$
$$m_{\tilde{\delta}}^2=(\rho_3/2-\rho_1)v_R^2- {\beta_1^2k_+^4k_0^2
\over{k_-^4(\alpha+\rho_1-\rho_3/2)}}.\eqno(25)$$
$$m_{\Delta_1}^2={\alpha_3k_-^2+4\rho_2v_R^2\over2}+
{k_-^4(\beta_3k_+^2+\beta_1k_1k_2)^2\over
{2k_1^4(4\rho_2 +\rho_3-2\rho_1)v_R^2}},\eqno(26)$$
$$m_{\Delta_2}^2={\alpha_3k_-^2-(2\rho_1-\rho_3)v_R^2\over2}-
{k_-^4(\beta_3k_+^2+\beta_1k_1k_2)^2\over
{2k_1^4(4\rho_2 +\rho_3-2\rho_1)v_R^2}},\eqno(27)$$
$$m_{S_1}^2=2\lambda_1k_+^2+8k_1^2k_2^2(2\lambda_2+\lambda_3)/k_+^2-
8\lambda_4k_1k_2+$$
$$+{4k_1k_2k_-^4[2(2\lambda_2+\lambda_3)k_1k_2/k_+^2-\lambda_4]^2
\over \alpha_2v_R^2k_+^2},\eqno(28)$$
where
$$\alpha={\alpha_3k_+^2\over2k_-^2}={\alpha_3(1+\tan^2\beta)\over2
(\tan^2\beta-1)},
\qquad \beta_0={\beta_1k_+^2\over k_-^2}={\beta_1(1+\tan^2\beta)\over
(\tan^2\beta-1)},$$
$$k_0={k_-^2\over \sqrt{2}k_+}={k_+(\tan^2\beta-1)\over\sqrt{2}(1+
\tan^2\beta)},$$
$\rho_{1,3}$ are the constants entering the
Yukawa potential, $\tan\beta=k_1/k_2$ and $v_R$ is the VEV of the neutral
component of the right-handed Higgs triplet, $v_R\gg\mbox{max}(k_1,k_2)$).
From the relations (25) and (27) it follows that the masses
$\tilde{\delta}^{(\pm)}$ and $\Delta^{(\pm\pm)}_2$-bosons are very close
to each other. For the $h^{(\pm)}$, $\tilde{\delta}^{(\pm)}$ and
$\Delta^{(\pm\pm)}_{1,2}$-boson masses to be around the electroweak scale
the constants $\alpha_3$, $\rho_2$ and $(\rho_3/2-\rho_1)$ should have the
order of $\sim10^{-2}$ which follows from the expressions (24) --- (27).

The Lagrangians which are required for our purposes are given by the
expressions
$${\cal{L}}_{\gamma\Delta\Delta}=2ie[(\partial_{\mu}\Delta^{(--)*}_1(x))
\Delta^{(--)}_1(x)-\Delta^{(--)*}_1(x)(\partial_{\mu}\Delta^{(--)}_1(x))]+
(1\rightarrow2),\eqno(29)$$
$${\cal {L}}^{dc}_l=-\sum_{a,b}{f_{ab}\over2}[\overline{l}^c_a(x)(1+
\gamma_5)l_b(x)c_{\theta_d}-\overline{l}^c_a(x)(1-\gamma_5)l_b(x)
s_{\theta_d}]\Delta^{(--)*}_1(x)+$$
$$+(1\rightarrow2, \theta_d\rightarrow\theta_d-{\pi\over2})+\mbox{conj}.,
\eqno(30)$$
$${\cal{L}}_{W_1\gamma h}={g_Rem_{W_1}(1-\tan^2\beta)(\alpha-\rho_3/2+
\rho_1+1)s_{\xi}\over g_L(1+\tan^2\beta)}h^{(-)*}(x)W_{1\mu}(x)A_{\mu}(x)
+\mbox{conj}.,\eqno(31)$$
$${\cal{L}}_{W_2\gamma h}={\cal{L}}_{W_1\gamma h}(s_{\xi}\rightarrow
c_{\xi}),\eqno(32)$$
$${\cal{L}}_{W_1\gamma \tilde{\delta}}=g_Reg_L^{-1}\beta_1m_{W_1}
s_{\xi}\tilde{\delta}^{(-)*}(x)W_{1\mu}(x)A_{\mu}(x)+\mbox{conj}.,\eqno(33)$$
$${\cal{L}}_{W_2\gamma\tilde{\delta}}={\cal{L}}_{W_1\gamma\tilde{\delta}}
(s_{\xi}\rightarrow c_{\xi}),\eqno(34)$$
$${\cal{L}}^{sc}_l=\sum_{a,b}\{[{h_{ab}^{\prime}k_2-h_{ab}k_1\over2k_+}
\overline{\nu}_a(x)(1-\gamma_5)l_b(x)-{h_{ab}k_2-h_{ab}^{\prime}k_1\over2k_+}
\overline{N}_a(x)(1+\gamma_5)l_b(x)]h^{(-)*}(x)+$$
$$+{f_{ab}\over{\sqrt{2}}}[\overline{l}^c_a(x)(1+\gamma_5)\nu_b(x)
\left({\beta_0
k^2_0\over{(\alpha+\rho_1-\rho_3/2)v_R^2}}h^{(-)*}(x)-\tilde{\delta}^{(-)*}(x)
\right)+\overline{l}^c_a(x)(1-$$
$$-\gamma_5)N_b(x)\left({k_0\over v_R}h^{(-)*}(x)+
{\beta_0k_0\over {(\alpha+\rho_1-
\rho_3/2)v_R}}\tilde{\delta}^{(-)*}(x)\right)]+\mbox{conj}.\},\eqno(35)$$
$${\cal{L}}^n=-{1\over\sqrt{2}k_+}\{\sum_{a,b}
\overline{l}_a(x)l_b(x)[(h_{ab}k_1+h^{\prime}_{ab}k_2)s_{\theta_0}
+(h^{\prime}_{ab}k_1-h_{ab}k_2)c_{\theta_0}]S_1(x),\eqno(36)$$
where the superscript $c$ denotes the charge conjugation operation,
$c_{\theta_d}=\cos\theta_d$, $s_{\theta_d}=\sin\theta_d$, $\theta_d$ is the
mixing angle of the doubly charged Higgs bosons
($\tan\theta_d\sim{k_+^2 /v_R^2}$), $f_{ab}$ is the Yukawa triplet coupling
constant, $g_R$ is the gauge coupling of the $SU(2)_R$ subgroup (further we
shall speculate that $g_L=g_R$), $N_a(x)$ describes the heavy neutrino with
the flavor $a$, $\xi$ is the mixing angle of the charged gauge bosons,
and the angle $\theta_0$
is determined by the Yukawa potential parameters and the VEV's
$$\tan2\theta_0={{4k_1k_2k_-^2[-2(2\lambda_2+\lambda_3)k_1k_2+\lambda_4k_+^2]}
\over{k_1k_2[(4\lambda_2+2\lambda_3)(k_-^4-4k_1^2k_2^2)-k_+^2
(2\lambda_1k_+^2-8\lambda_4k_1k_2)]+\alpha_2v_R^2k_+^4}}.\eqno(37)$$

The influence of the Yukawa potential choice on the physical results
could be easily seen by the example of the Lagrangian (36). Really, when
in the condition (23) the change
$k_2\rightarrow-k_2$ is carried out, then instead of (36) one obtains
$${\cal{L}}^n=-{1\over\sqrt{2}k_+}\{\sum_am_a\overline{l}_a(x)l_a(x)
c_{\theta_0}+\sum_{a,b}\overline{l}_a(x)l_b(x)(h_{ab}k_1-
h^{\prime}_{ab}k_2)s_{\theta_0}\}S_1(x).\eqno(38)$$
Since $\tan\beta_0\sim k_-^2/v_R^2$ and for the muon $m_{\mu}/k_+\sim6
\times10^{-3}$, then, as the exapmle, the cross section of the electron-muon
recharge
$$e^-\mu^+\rightarrow e^+\mu^-,$$
never could have the resonance peak connected with the $S_1$-boson when
the Lagrangian (38) is used [39], while the existence of such a peak could be
quite possible when one works with the Lagrangian (36) [40].

We now proceed to the calculations of the contribution to the muon
AMM caused by the Higgs bosons. The diagrams corresponding the
exchange of the doubly charged Higgs bosons are shown in Fig.1.
They give the following corrections to the AMM value
$${\delta a_{\mu}^{\Delta}\over\mu_0}={1\over8\pi^2}\left(
4f_{\mu e}^2\sum_{i=1}^2I^{\Delta_i}_e+
f_{\mu\mu}^2\sum_{i=1}^2I^{\Delta_i}_{\mu}+
4f_{\mu\tau}^2\sum_{i=1}^2I^{\Delta_i}_{\tau}\right),\eqno(39)$$
where
$$I^{\Delta_i}_{l_a}=\int^1_0\left({2m_{\mu}^2(z^2-z^3)\over m_{\mu}^2(z^2-z)+
m_{\Delta_i}^2z+m_{l_a}^2(1-z)}+{m_{\mu}^2(z^2-z^3)\over m_{\mu}^2(z^2-z)+
m_{\Delta_i}^2(1-z)+m_{l_a}^2z}\right)dz,$$
and $I^{\Delta_i}_{l_a}>0.$

The singly charged Higgs bosons also influence the value of the AMM.
The relevant diagrams are depicted in Fig.2. For the diagrams which contain
the loops with the $W_1^{\pm}-$ and $h^{(\pm)}$-bosons the following relation
takes place
$${M_{W_1N_{\mu}h}\over M_{W_1\nu_{\mu}h}}=s_{\xi},$$
where $M_{W_1\nu_{\mu}h}$ ($M_{W_1N_{\mu}h}$) are the matrix elements
appropriate the diagrams with the exchange of the light (heavy) neutrino.
As the mixing angle of the charged gauge bosons is very small
$\mid\xi\mid \approx10^{-2}-10^{-5}$ [33], then
one could neglect the contributions coming from the diagrams with the virtual
heavy neutrino. Taking into account the analogous relations
$${M_{W_2\nu_{\mu}h}\over M_{W_2N_{\mu}h}}=s_{\xi},\qquad
{M_{W_1N_{\mu}\tilde{\delta}}\over M_{W_1\nu_{\mu}\tilde{\delta}}}=s_{\xi},
\qquad {M_{W_2\nu_{\mu}\tilde{\delta}}\over M_{W_2
N_{\mu}\tilde{\delta}}}=s_{\xi},$$
we give the dominant contribution from the diagrams shown in Fig.2 to the muon
AMM
$${\delta a_{\mu}^{(hh)}\over\mu_0}={1\over8\pi^2}
\sum_{a=e,\mu,\tau}\left(\alpha_{\mu N_ah}^2I^{hh}_{N_a}+\alpha_{\mu\nu_ah}^2I^{hh}_{\nu_a}
\right),\eqno(40)$$
$${\delta a_{\mu}^{(\tilde{\delta}\tilde{\delta})}\over\mu_0}=
{1\over8\pi^2}\sum_{a=e,\mu,\tau}\left(\alpha_{\mu N_a\tilde{\delta}}^2
I^{\tilde{\delta}\tilde{\delta}}_{N_a}+\alpha_{\mu\nu_a\tilde{\delta}}^2
I^{\tilde{\delta}\tilde{\delta}}_{\nu_a}\right),\eqno(41)$$
$${\delta a_{\mu}^{(W_1h)}\over\mu_0}=
{(\alpha-\rho_3/2+\rho_1+1)(1-\tan^2\beta)s_{\xi}m_{W_1}\over
16\sqrt{2}\pi^2(1+\tan^2\beta)}\alpha_{\mu\nu_{\mu}h}I^{W_1h},\eqno(42)$$
$${\delta a_{\mu}^{(W_2h)}\over\mu_0}=
{(\alpha-\rho_3/2+\rho_1+1)(1-\tan^2\beta)c_{\xi}m_{W_1}\over16\sqrt{2}
\pi^2(1+\tan^2\beta)}\alpha_{\mu N_{\mu}h}I^{W_2h},\eqno(43)$$
$${\delta a_{\mu}^{(W_1\tilde{\delta})}\over\mu_0}={\beta_1
(1-\tan^2\beta)s_{\xi}m_{W_1}\alpha_{\mu\nu_{\mu}\tilde{\delta}}\over16
\sqrt{2}\pi^2(1+\tan^2\beta)}I^{W_1\tilde{\delta}},\eqno(44)$$
$${\delta a_{\mu}^{(W_2\tilde{\delta})}\over\mu_0}=
{\beta_1(1-\tan^2\beta)c_{\xi}m_{W_1}\alpha_{\mu N_{\mu}\tilde{\delta}}\over16
\sqrt{2}\pi^2(1+\tan^2\beta)}I^{W_2\tilde{\delta}},\eqno(45)$$
where
$$\alpha_{l_a\nu_bh}={h_{ab}^{\prime}k_2-h_{ab}k_1\over2k_+},\qquad
\alpha_{l_aN_bh}={h_{ab}^{\prime}k_1-h_{ab}k_2\over2k_+},\qquad
\alpha_{l_a\nu_b\tilde{\delta}}={f_{ab}\over{\sqrt{2}}},$$
$$\alpha_{l_aN_b\tilde{\delta}}={f_{ab}\beta_0k_0\over\sqrt{2}
(\alpha+\rho_1-\rho_3/2)v_R}={f_{ab}\beta_1k_+\over2(\alpha+\rho_1-
\rho_3/2)v_R},$$
$$I^{hh}_i=\int^1_0
{m_{\mu}^2(z^3-z^2)dz\over m_{\mu}^2z^2+(m_h^2-m_i^2-m_{\mu}^2)z+
m_i^2},\qquad i=\nu_a, N_a,$$
$$I^{\tilde{\delta}\tilde{\delta}}_i=
I^{hh}_i(m_h\rightarrow m_{\tilde{\delta}}), \qquad I^{hh}_i<0$$
$$I^{W_1h}={m_{\mu}\over m_{W_1}^2-m_h^2}
\{\ln\left({m_{W_1}^2\over m_h^2}\right)
-\int_0^1{z^2[m_{\mu}^2(2z-1)+m_{W_1}^2-m_{\nu_{\mu}}^2]dz
\over m_{\mu}^2z^2+(m_{W_1}^2-m_{\nu_{\mu}}^2-m_{\mu}^2)z+
m_{\nu_{\mu}}^2}+(m_{W_1}\rightarrow m_h)\},$$
$$I^{W_2h}=I^{W_1h}(m_{W_1}\rightarrow m_{W_2}, m_{\nu_{\mu}}\rightarrow
m_{N_{\mu}}),\qquad
I^{W_k\tilde{\delta}}=I^{W_kh}(m_h\rightarrow m_{\tilde{\delta}}),
\qquad I^{W_kh}>0.$$

The contribution from the neutral Higgs boson $S_1$ to the AMM value
are due to the diagram shown in Fig.3 and is given by
$${\delta a_{\mu}^{(S_1)}\over\mu_0}={1\over8\pi^2}\sum_a\alpha_{\mu l_aS_1}^2
I^{S_1}_{l_a},\eqno(46)$$
where
$$\alpha_{l_al_bS_1}=-{1\over\sqrt{2}k_+}
[(h_{ab}k_1+h^{\prime}_{ab}k_2)s_{\theta_0}+
(h^{\prime}_{ab}k_1-h_{ab}k_2)c_{\theta_0}],$$
$$I^{S_1}_{l_a}=\int^1_0{[m_{\mu}^2(z^2-z^3)+m_{l_a}^2z^2]dz\over m_{\mu}^2
(z^2-z)+m_{S_1}^2(1-z)+m_{l_a}^2z},\qquad I^{S_1}_{l_a}>0.$$

The total correction value motivated by the Higs bosons to the
muon AMM $\delta a_{\mu}/\mu_0$ is defined by the sum of the
expressions (39) --- (46). To perform an exhaustive analysis of
the obtained result one should have information both about the
coupling constants $\alpha_{L_aL_bH_i}$  ($L_a=l_a,\nu_a,N_a$) and
the Higgs boson masses $m_{H_i}$. At present such information
follows from looking for the deviations from the SM predictions.
It is usually reported in terms of the upper limits for quantities
of the type $\alpha_{L_aL_bH_i}/m_{H_i}$ or, that is more
frequent, for quantities of the type
$$\sum_iC_i\epsilon_i^{aba^{\prime}b^{\prime}}=\sum_iC_i{(\alpha_{L_aL_bH_i}
\alpha_{L_{a^{\prime}}L_{b^{\prime}}H_i})^2\over m_{H_i}^4},$$
where $C_i$ are the constants (see, for review [41]). As a rule
the determination of the upper bound only for one quantity
$\alpha_{L_aL_bH_i}/m_{H_i}$ is a very involved task. As a case in
point we consider the decay $$\mu^-\rightarrow
e^-\gamma.\eqno(47)$$ In the third order of the perturbation
theory the appropriate diagrams follow from those in Figs.1,2 and
3 when in the final state the muon is replaced by the electron.
The width of the decay (47) includes the whole complex quantities
$\epsilon_i^{aba^{\prime}b^{\prime}}$. Besides, the presence of
the diagrams with the singly charged Higgs bosons leads to the
destructive interference of the reaction amplitudes that
complicates a lot the task of extracting information about the
individual quantities of $\epsilon_i^{aba^{\prime}b^{\prime}}$.

However, the Higgs bosons coupling constants $\alpha_{\overline{L}_aL_bH_i}$
could be expressed in terms of the lepton sector parameters. It is easily done
at least in the two-flavor approximation.
For this purpose we need the neutrino mass matrix ${\cal M}$.
Once one chooses the basis
$\Psi^T=\left(\nu_{aL}^T,N_{aR}^T,\nu_{bL}^T,N_{bR}^T\right)$,  the
${\cal M}$ takes the form
$${\cal M}=\left(\matrix{f_{aa}v_L & m^a_D & f_{ab}v_L & M_D\cr
m_D^a & f_{aa}v_R & M_D & f_{ab}v_R\cr
f_{ab}v_L & M_D & f_{bb}v_L & m^b_D \cr
M_D & f_{ab}v_R & m_D^b & f_{bb}v_R\cr}\right),\eqno(48)$$
where $v_L $ is the VEV of the neutral component of the left-handed
Higgs triplet ($v_L\ll(\mbox{max}(k_1,k_2)$) and
$$m_D^a=h_{aa}k_1+h^{\prime}_{aa}k_2,\eqno(49)$$
$$M_D=h_{ab}k_1+h^{\prime}_{ab}k_2.\eqno(50)$$

In its turn the elements of the matrix ${\cal M}$ are connected with the
neutrino oscillations parameters [39,42]
$$\left.\begin{array}{lll}
m_D^a=c_{\varphi_a}s_{\varphi_a}(-m_1c^2_{\theta_{\nu}}-m_3s^2_
{\theta_{\nu}}+m_2c^2_{\theta_N}+m_4s^2_{\theta_N}),\\
\\
m_D^b=m_D^a(\varphi_a\rightarrow \varphi_b,\theta_{\nu,N}
\rightarrow\theta_{\nu,N}+{\pi\over2}),\end{array}\right\}\eqno(51)$$
$$M_D=c_{\varphi_a}s_{\varphi_b}c_{\theta_{\nu}}s_{\theta_{\nu}}
(m_1-m_3)+s_{\varphi_a}c_{\varphi_b}c_{\theta_N}s_{\theta_N}(m_4-m_2),
\eqno(52)$$
$$f_{ab}v_R=s_{\varphi_a}s_{\varphi_b}c_{\theta_{\nu}}s_{\theta_{\nu}}
(m_3-m_1)+c_{\varphi_a}c_{\varphi_b}c_{\theta_N}s_{\theta_N}(m_4-m_2),\eqno(53)$$
$$\left.\begin{array}{lll}
f_{aa}v_R=(s_{\varphi_a}c_{\theta_{\nu}})^2m_1+(c_{\varphi_a}c_{\theta_N})
^2m_2+(s_{\varphi_a}s_{\theta_{\nu}})^2m_3+(c_{\varphi_a}s_{\theta_N})^2m_4,\\
\\
f_{bb}v_R=f_{aa}v_R(\varphi_a\rightarrow\varphi_b+{\pi\over2},
\theta_N\rightarrow\theta_N+{\pi\over2}),\end{array}\right\}\eqno(54)$$
$$f_{ll^{\prime}}v_L=f_{ll^{\prime}}v_R(\varphi_{l,l^{\prime}}\rightarrow
\varphi_{l,l^{\prime}}+{\pi\over2}),\qquad l,l^{\prime}=a,b,\eqno(55)$$
where $\varphi_a$ is the mixing angle in the $a$ generation between the
light and the heavy neutrino entering into the left-handed and the right-handed
lepton doublet
$$\left(\matrix{\nu_a\cr l_a}\right)_L,\qquad
\left(\matrix{N_a\cr l_a}\right)_R,$$
respectively,
$\theta_{\nu} (\theta_N)$ is the mixing angle between the $\nu_{aL}$ and the
$\nu_{bL}$ neutrino ($N_{aR}$ and $N_{bR}$), $c_{\varphi_a}=
\cos\varphi_a, \ s_{\varphi_a}=\sin\varphi_a$ and so on.
As $m_{\nu}\ll m_N$, then with the help of Eqs.(53) and (55) it is possible
to find the relationship for an estimation of the mixing angles between light
and heavy neutrinos. Further on we shall assume that the mixing takes place
between $\mu$ and $\tau$ generations ($a=\mu, b=\tau$) only. Then for the mixing
angles we obtain
$$\sin2\varphi_{\mu}\approx{f_{\mu\mu}\sqrt{v_Rv_L}\over c_{\theta_N}^2
m_{N_{\mu}}+s_{\theta_N}^2m_{N_{\tau}}},\eqno(56)$$
$$\sin2\varphi_{\tau}\approx{f_{\tau\tau}\sqrt{v_Rv_L}\over s_{\theta_N}^2
m_{N_{\mu}}+c_{\theta_N}^2m_{N_{\tau}}}.\eqno(57)$$
The estimation of $v_L $ could be done with the help of quantity
$$\rho={m_Z^2c_W^2\over m_W^2}.$$
In LRM the quantity $\rho$ is defined by the relation [43]
$$\rho={1+4x\over1+2x},\eqno(58)$$
where
$$x=\left({v_L\over k_+}\right)^2.$$
As the experiment for today yields
$$\rho=1.0107\pm0.0006, $$
that the value $v_L $ can reach 13 GeV.

Taking into consideration both the definition of $m^a_D$ (Eqs. (49), (51) )
and the formulae for the charged lepton masses
$$m_{l_a}=h_{aa}k_2+h_{aa}^{\prime}k_1\eqno(59)$$
it is not difficult to obtain
$$\alpha_{\overline{l}_a\nu_ah}={h_{aa}^{\prime}k_2-h_{aa}k_1\over2k_+}=
{1+\tan^2\beta\over2k_+(1-\tan^2\beta)}\left({2m_{l_a}\tan\beta\over
1+\tan^2\beta}+m^a_D\right)\approx$$
$$\approx{1+\tan^2\beta\over2k_+(1-\tan^2\beta)}\left[{2m_{l_a}\tan\beta\over
1+\tan^2\beta}+c_{\varphi_a}s_{\varphi_a}(m_2c^2_{\theta_N}+m_4s^2_{\theta_N})
\right].\eqno(60)$$
The analogous mathematics for $\alpha_{\overline{l}_aN_ah}$ lead
to the expression
$$\alpha_{\overline{l}_aN_ah}={h_{aa}^{\prime}k_1-h_{aa}k_2\over2k_+}=
{1+\tan^2\beta\over2k_+(1-\tan^2\beta)}\left({2m^a_D\tan\beta\over
1+\tan^2\beta}-m_{l_a}\right)\approx$$
$$\approx{1+\tan^2\beta\over2k_+(1-\tan^2\beta)}\left[{2c_{\varphi_a}
s_{\varphi_a}(m_2c^2_{\theta_N}+m_4s^2_{\theta_N})\tan\beta\over1+\tan^2\beta}-
m_{l_a}\right].\eqno(61)$$
It is pertinent to note that there is the connection between
the coupling constants $\alpha_{\overline{l}_aN_bh}$ and
$\alpha_{\overline{l}_al_bS_1}$
$$\alpha_{\overline{l}_aN_bh}\approx{\alpha_{\overline{l}_al_bS_1}
\over\sqrt{2}}.\eqno(62)$$

The next step is the determination of the non-diagonal Higgs bosons
coupling constants. To suppress the mixing in the charged lepton sector
(betwen $l_a$ and $l_b$) it is necessary to demand
$$h_{ab}k_2+h_{ab}^{\prime}k_1=0.\eqno(63)$$
Then, with the regard to the definitions of the quantity $M_D$ (Eqs.(45)
and (47) ) one obtains
$$\alpha_{\overline{l}_a\nu_bh}=-{M_D\over2k_+}
\approx-{s_{\varphi_a}c_{\varphi_b}c_{\theta_N}s_{\theta_N}(m_4-m_2)\over2k_+},
\eqno(64)$$
$$\alpha_{\overline{l}_aN_bh}=
-{M_D\tan\beta\over k_+(1+\tan^2\beta)}\approx
-{s_{\varphi_a}c_{\varphi_b}c_{\theta_N}s_{\theta_N}(m_4-m_2)\tan\beta\over
k_+(1+\tan^2\beta)}.\eqno(65)$$

From the expressions (60) --- (62), (64) and (65) it is obvious
that the values of the coupling constants
$\alpha_{\overline{L}_aL_bH_i}$ are basically defined by the
oscillation parameters of the heavy neutrinos. Nowadays the
information concerning the heavy neutrino sector is very poor. All
we have is the upper bound for the heavy electron neutrino mass
resulting from the experiments aimed at finding the neutrinoless
double $\beta$ decay $$m_{N_e}>63\ \mbox{GeV}\left({1.6\
\mbox{TeV}\over m_{W_2}}\right)^4. \eqno(66)$$ Hence, our sole way
out in an existing situation is to set any minimal number of
parameters of heavy neutrinos sector, and other parameters to
express through them with the help of the equations (51) --- (55).
As those we shall take $m_{N_{\mu}}$, $m_{N_{\tau}}$ and
$\theta_N$.

Now we shall pass to the discussion of the constraints on the Higgs bosons
masses. The current limit on the singly charged Higgs boson mass
has been obtained within the THDM's under investigation of the reaction
$$e^+e^-\rightarrow H^+H^-.\eqno(67)$$
The lowest value for the mass of the charged Higgs boson,
independent of the its branching ratio, is currently 78.6 GeV [33]. It is
evident that this limit may be broken down to the singly charged Higgs bosons
of the LRM $\tilde{\delta}^{(\pm)}$ and $h^{(\pm)}$. Really,
in the THDM the charged Higgs boson interacts with the quarks at the tree
level while in the LRM such interaction exists for the $h^{(\pm)}$-boson
only.
Furthermore, in the both models the coupling constants of the charged Higgs
bosons with the $Z$-boson are not equal to each other
$${(g_{HHZ})_{2HDM}\over (g_{\tilde{\delta}\tilde{\delta}Z})_{LRM}}=
{\cot2\theta_W\over g^{\prime}\cos^{-1}\theta_W(\alpha+\rho_1-\rho_3/2)
(g^{\prime-1}\sin\theta_W\cos\Phi+g_R^{-1}\sin\Phi)},\eqno(68)$$
$${(g_{HHZ})_{2HDM}\over (g_{hhZ})_{LRM}}=-
{\cot2\theta_W\over\cos^{-1}\theta_W(\sin^{-1}\theta_W\cos2\theta_W\cos\Phi+g_R
g^{\prime-1}\sin\Phi)/2}.\eqno(69)$$
However, since the analysis of the process (67) from the LRM poit of view
is absent up to now, we shall assume that the lower bound on the masses of
the singly charged Higgs bosons of the LRM is 78.6 GeV too.

As regards the doubly charged Higgs boson mass the situation is somewhat
more simple. The $\Delta^{(\pm\pm)}_{1,2}$-bosons are typical
representatives of the LRM. For this reason the experiments aimed at their
appearence are
analyzed from the LRM poit of view only. The current lower bounds on their
masses
obtained by OPAL Collaboration at 95\% $CL$ [44] are 98.5 GeV.

We also should discuss the implementation of the lower bound 115 GeV
on the mass of the lightest neutral Higgs (LNH) in the SM extensions.
This bound has been obtained on LEP II under investigation of the reaction
$$e^+e^-\rightarrow Z^*\rightarrow Zh\eqno(70)$$
from the SM poit of view [45]. The reaction (70) is analyzed for the four
$Zh$ decay channels
$$Zh\rightarrow q\overline{q}q^{\prime}\overline{q}^{\prime},
q\overline{q}\nu\overline{\nu}, q\overline{q}l_a\overline{l}_a\ (l_a=e,\mu),
\tau^+\tau^-q\overline{q},$$
where the final state $h\rightarrow q\overline{q}$ includes both
the quark-antiquark and the gluon-gluon pairs. For the THDM's, as an example,
the substantial deviations from the SM are present both in the cross section
$\sigma_{e^+e^-\rightarrow Zh}$ and in the decay widths $\Gamma_h$.
Since in these models the coupling constant of the neutral CP-even
$h$ boson (analog of the SM Higgs boson) with the $Z$ boson has the form
$$g_{ZZh}={g_Lm_Z\sin(\beta-\alpha)\over\mbox{c}_{\theta_W}}\eqno(71)$$
where $\sin(\beta-\alpha)\sim0$,
then $\left(\sigma_{e^+e^-\rightarrow Zh}\right)_{THDM}$
is much less relatively to the SM value. On the other hand,
as the relation takes place
$${\left(g_{f\overline{f}h}\right)_{THDM}\over\left(g_{f\overline{f}h}
\right)_{SM}}\approx\tan\beta,\qquad f=b,\tau.$$
the $h$ boson decay widths through quarks and leptons have
greater values than in the SM. It is evident, that the analysis of the reaction
(70) leads to the different mass values for the SM Higgs and the LNH of the
THDM. Actually, when
$\mid\sin(\beta-\alpha)\mid\leq0.06$ the LEP data result in $m_h\sim10$ GeV
at 98\% CL [46].

The only reason for the LRM cross section $\sigma_{e^+e^-\rightarrow ZS_1}$
not to coincide with that of the SM  may be the coupling constant
of the $S_1$ boson with the $Z_1$ boson
$$g_{Z_1Z_1S_1}={g_L\mbox{s}_W^2m_{Z_1}[\mbox{c}_{\theta_0}(1-\mbox{tg}^2\beta)-
2\mbox{s}_{\theta_0}\tan\beta)]\over 2\mbox{c}_W(\tan^2\beta+1)}
\left(2g_Rg^{\prime-1}\mbox{c}_{\Phi}\mbox{s}_{\Phi}\mbox{s}_W^{-1}-\right.$$
$$\left.-\mbox{c}_{\Phi}^2\mbox{s}_W^{-2}-g_R^2g^{\prime-2}\mbox{s}_{\Phi}^2
\right),\eqno(72)$$
where $\Phi$ is the mixing angle of $Z_1$ and $Z_2$ bosons ($\Phi\approx
10^{-2}-10^{-3}$) and $g^{\prime}$ is the gauge constant of subgroup
$U(1)_{B-L}$. When
$$\Phi=k_2=0$$
then $g_{Z_1Z_1S_1}$ converts to the constant describing the interaction
between the Higgs bosons and the $Z$ boson in the SM.
Since the symmetric LRM reproduces the SM under the following values of
$g_R$ and $g^{\prime}$
$$g_L=g_R=e\mbox{s}_W^{-1},\qquad  g^{\prime}=e\sqrt{\mbox{c}_W^2-
\mbox{s}_W^2},\eqno(73)$$
then the quantity $g_Rg^{\prime-1}$ may moderately differ from unity.
Therefore, the major contributor to the deviation $g_{Z_1Z_1S_1}$
from their SM values is the factor
$$\Delta g={[\mbox{c}_{\theta_0}(1-\tan^2\beta)-
2\mbox{s}_{\theta_0}\tan\beta)]\over (\tan^2\beta+1)}.\eqno(74)$$
From the expression (37) follows that the angle value $\theta_0$
is basically determined by the parameter $\alpha_2$ which enters the Higgs
potential. When $\alpha_2\sim10^{-2}$
then the angle $\theta_0$ may reach the value $\pi/4$. At this condition
the $S_1$ boson could remain light as usual but the $S_2$ boson
ceases to be superheavy
$$m_{S_2}^2={\alpha_2v_R^2k_+^2\over k_1k_2}-
{4k_1k_2k_-^4[2(2\lambda_2+\lambda_3)k_1k_2/k_+^2-\lambda_4]^2
\over \alpha_2v_R^2k_+^2}.\eqno(75)$$
Recall, that the demand
$$m_{S_2}\geq10\ \mbox{TeV},\eqno(76)$$
is caused by the necessity to supress at the tree level the flavor changing
neutral currents (FCNC) in the Lagrangian
$${\cal L}^n_q=-{1\over\sqrt{2}k_+}\sum_{a,b}\overline{u}_a\left\{\left[
m_{u_a}\left(\mbox{c}_{\theta_0}+{2k_1k_2\over k_-^2}\mbox{s}_{\theta_0}
\right)S_1-m_{u_a}(\mbox{s}_{\theta_0}-{2k_1k_2\over k_-^2}
\mbox{c}_{\theta_0})S_2-\right.\right. $$
$$\left.\left.-im_{d_a}\gamma_5P_1\right]\delta_{ab}+{k_+^2\over k_-^2}
\left({\cal K}{\cal M}_d{\cal K}^*\right)_{ab}\left(S_1\mbox{s}_{\theta_0}+
S_2\mbox{c}_{\theta_0}\right)\right\}u_b+$$
$$+(u_a\rightarrow d_a, m_{u_a}\leftrightarrow m_{d_a}, \gamma_5\rightarrow-
\gamma_5),\eqno(77)$$
where ${\cal{K}}$ is the Cabibbo-Kobayashi-Maskawa matrix and ${\cal{M}}_d$ is
the diagonal mass matrix for the down quarks.
The absence of the FCNC in its turn allows to discribe properly
the $\overline{K}^0\leftrightarrow K^0$ transitions.
However, as it is shown in [39] the successfull LRM building
demands the redefinition of the traditional Yukawa Lagrangian
for quarks. The expression (76) must be changed for
$${\cal L}^n_q=-{1\over \sqrt{2}k_+}\sum_a \overline{u}_a\left\{m_{u_a}\left[
\left(\mbox{c}_{\theta_0}-{k_1\over k_2}\mbox{s}_{\theta_0}\right)S_1-
\left(\mbox{s}_{\theta_0}+{k_1\over k_2}\mbox{c}_{\theta_0}\right)S_2\right]+
\right.$$
$$\left.+{im_{u_a}k_1\over k_2}\gamma_5P_1\right\}u_a+
(u_a\rightarrow d_a, \theta_0\rightarrow-\theta_0).\eqno(78)$$
Since the Lagrangian (78) does not induce any FCNC, the inequality
(76) breaks down. Then from Eq.(74) follows that with the increasing of the
angle
$\theta_0$ the deviation $\Delta g$ from unity could be large enough.

From the form of the Lagrangian (78) it is evident that the decay widths
of the $S_1$ boson into quarks and gluons may significantly differ from
those of the SM. Since the coupling constant of the $S_1$ boson with the
$\tau$ lepton is determined by Eq.(62) then the value
$\Gamma_{S_1\rightarrow\tau^+\tau^-}$ also could not  coincide
with the corresponding value in the SM.
Thus, it is apparent that the LNH mass lower bound in the LRM may not
agree with that in the SM.
However, since up to now any works containing the analysis of the process (70)
from the point of view of the LRM are absent, we shall also take the
value $115$ GeV as the low bound on the $S_1$-boson mass.

Now we are ready to embark on the investigation of the Higgs boson contributions
to the muon AMM. Let us determine some key moments in our strategy at
calculation of the couplings constants of the Higgs bosons.
To evaluate the VEV of the right-handed Higgs triplet $v_R$ we invoke the
relation [39]
$$v_R=\sqrt{{m_{W_2}^2-m_{W_1}^2\over g_L^2(1+\mbox{tg}^22\xi)}}.\eqno(79)$$
The current bounds on the $W_2$ gauge boson and the mixing angle
$\xi$ are varied within a broad range in relation to
what kind of reactions and what assumptions have been used at analysis [33].
For example, the lower bound on $m_{W_2}$ being equal to 484 GeV is obtained
from the investigation of the polarized muon decay under assumption $\xi=0$.
The analysis of the process $b\rightarrow s\gamma$ leads to the
constraints
$$-0.01\leq\xi\leq0.003.\eqno(80)$$
Having specified $m_{W_2}=0.8$ TeV and $\xi=10^{-2}$ we evaluate $v_R$.
Then setting values $m_{N_{\mu}}$, $m_{N_{\tau}}$, $\theta_N$ and $v_L$ we
can present the quantities $\varphi_{\mu}$, $\varphi_{\tau}$,
$f_{\mu\tau}$, $m^{\mu}_D$ and $M_D$ as the functions on $f_{\mu\mu}$.

First we assume that the dominant contribution comes from
the $\Delta^{(--)}_2$-boson. A negligible value of the corrections
from the $S_1$, $\Delta^{(--)}_1$, $\tilde{\delta}^{(-)}$ and $h^{(-)}$-bosons
could be caused both by the large values of their masses and by
the small values of their couplings constants.
It is natural to require that $f_{\mu\mu} $ should be less than 1. Then
analysis shows that the interval of the $\Delta_2^{(--)}$-boson mass at which
the satisfaction to BNL'00-results is possible,
critically depends on the value of $f_{\mu\tau}$.
Let us note, that the value of $f_{\mu\tau}$, as follows from Eq. (53),
very weakly depends on the angles $\varphi_{\mu}$ and $\varphi_{\tau}$
and is basicly determined by the difference of the heavy neutrino masses.
When one sets $v_L$ equal to 1.7 GeV then the $\Delta_2^{(--)}$-boson mass
would reach the greatest value $(m_{\Delta_2})_{max}$ at $f_{\mu\tau}\approx
0.15$.  For this case in the $m_{\Delta_2}$ {\it{vs.}} $f_{\mu\mu}$ parameter
space two contour lines marked  $93.8$ and $512.8$
corresponding to $95\%$ CL limits for the contribution of New Physics
to $\delta a_{\mu}/\mu_0$ are exhibited in Fig.4.
The range of the Higgs sector parameters allowed by the BNL'00 result
lies between contours $93.8$ and $512.8$.

When $f_{\mu\tau}>0.15$ and $m_{\Delta_2}>140$ GeV the value of $f_{\mu\mu}$
becomes more than 1 for the upper bound of $\delta a_{\mu}/\mu_0$.
Decreasing of $f_{\mu\tau}$ results in the reduction of $(m_{\Delta_2})_{max}$.
At fixed $f_{\mu\tau}$ the reduction of $v_L$ practically has
no effect on the final result. However, increasing of $v_L$ up to 10 GeV
gives rise to the growth of $f_{\mu\mu}$. For example, at $m_{\Delta_2}
=100$ GeV the value of $f_{\mu\mu}$ lies in the interval $(0.318,0.743)$.

Inasmuch as the masses of the $\Delta^{(--)}_1$ and $\tilde{\delta}^{(-)}$-
bosons are close
to each other then the following possibility should be considered:
the observable value of the muon AMM stems from $\Delta^{(--)}_{1,2}$ and
$\tilde{\delta}^{(-)}$-bosons. Notice, that the quantities
$\delta a_{\mu}^{(W_{1,2}\tilde{\delta})}$ change the sign as one passes from
the region $\tan\beta<1$ to the region $\tan\beta>1$. We shall
restrict our consideration to the specific case, namely, when the quantities
$\beta_1\alpha_{\mu\nu_{\mu}\tilde{\delta}}$ and
$\beta_1\alpha_{\mu N_{\mu}\tilde{\delta}}$  are positive (recall, that
$I^{(W_{1,2}\tilde{\delta})}>0$).

In numerical calculations we shall use the following parameters values
$$\left.\begin{array}{ll}
m_{N_{\mu}}=110\ \mbox{GeV},\qquad m_{N_{\tau}}=125\ \mbox{GeV},
\qquad v_L=0.17\ \mbox{GeV},\\
\tan\beta=0.8,\qquad\alpha-\rho_3/2+\rho_1=1,\qquad \beta_1=1,\qquad
\theta_N=0.78\end{array}\right\}\eqno(81)$$
and shall assume such hierarchy of the Higgs boson masses
$$m_{\Delta_1}=1.1\ m_{\Delta_2},\qquad m_{\tilde{\delta}}=1.05\ m_{\Delta_2}.$$
In the $m_{\Delta_2}$ {\it{vs.}} $f_{\mu\mu}$ parameter
space two contour lines marked  $93.8$ and $512.8$
are shown in Fig.5. At the chosen values of the heavy neutrino masses the
quantity $f_{\mu\tau}$ is approximately equal to 0.01.

With the increase of the heavy neutrino masses, but provided that
$$m_{N_{\tau}}-m_{N_{\mu}}=\mbox{const}$$
the function $f_{\mu\mu}(m_{\Delta_2})$ grows faster.
However, this rise becomes essential only when the heavy neutrino masses
are approximately changed on the order of magnitude.
So, for example, choosing $m_{N_{\mu}}=900$ and $m_{N_{\tau}}=915$ GeV,
we obtain that at $m_{\Delta_2}=200$ GeV the value of $f_{\mu\mu}$ lies in
the interval $(0.092, 0.444)$. In this case $f_{\mu\tau}$ is approximately
equal to 0.006. On the other hand, with the increase of
$m_{N_{\tau}}-m_{N_{\mu}}$ the rate of growth of the function
$f_{\mu\mu}(m_{\Delta_2})$ goes down. For example, when we set
$$m_{N_{\mu}}=900\ \mbox {GeV}, \qquad m_{N_{\tau}}=1100\ \mbox{GeV}$$
and leave all the remaining parameters without change, the value of
$f_{\mu\mu}$
will lie in the interval $(0.029, 0.235)$ at $m_{\Delta_2}$ being equal to
100 GeV. The reduction of $\tan\beta$ results in decreasing of $f_{\mu\mu}$ as
a function of $m_{\Delta_2}$. For example, in the case $m_{N_{\mu}}=110$
and $m_{N_{\tau}}=125$ GeV (all the remaining parameter values are unchanged)
when $\tan\beta$ has been set to 0.3 we have
$$f_{\mu\mu}\in(0.009,0.049)\qquad\mbox{when}\qquad m_{\Delta_2}=100\
\mbox{GeV}$$
and
$$f_{\mu\mu}\in(0.014, 0.079) \qquad\mbox{when}\qquad m_{\Delta_2}=200\
\mbox{GeV}.$$
The obtained results are practically not changed at increasing of $v_L$
up to its maximum value.

At present we assume that the muon AMM value could be explained
by the $S_1$-boson contribution only. To suppress the contributions coming from
the remaining Higgs bosons it is enough to assume that
$$\alpha\sim1, \qquad\rho_2\sim1, \qquad \rho_3/2-\rho_1\sim1\eqno(82)$$
(this will make $h^{(-)}$-, $\Delta^{(--)}_{1,2}-$ and $\tilde{\delta}^{(-)}$-
bosons to be superheavy). The contours $93.8$ and $512.8$ in $m_{S_1}$
{\it{vs.}} $\tan\beta$ parameter space are represented in Fig.6.
In numerical calculations the following parameters values have been used
$$f_{\mu\mu}=0.04, \qquad m_{N_{\mu}}=110\ \mbox{GeV},\qquad m_{N_{\tau}}=125\
\mbox{GeV},\qquad v_L=1.7\ \mbox{GeV}.$$
Increasing (decreasing) of $v_L$ results in the reduction
(the enhancement) of the allowed values of $\tan\beta$. For example, when
$v_L$ is equal to 10 GeV we obtain
$$\tan\beta\in(0.804, 0.913)\qquad\mbox{at}\qquad m_{S_1}=115\ \mbox{GeV}$$
and
$$\tan\beta\in(0.88, 0.947)\qquad\mbox{at}\qquad m_{S_1}=200\ \mbox{GeV}.$$
Increasing of the heavy neutrino masses and of the values of $f_{\mu\mu}$
does not cause the appreciable change of the obtained results.

In the case when the muon €ŒŒ is stipulated by the contributions from
$S_1$- and $h$-bosons the contours $93.8$ and $512.8$ in $m_{S_1}$
{\it{vs.}} $\tan\beta$ parameter space are represented in Fig.7.
The choice of the model parameters is as follows
$$m_{N_{\mu}}=110\ \mbox{GeV},\qquad m_{N_{\tau}}=125\ \mbox{GeV},\qquad
v_L=1.7\ \mbox{GeV},\qquad f_{\mu\mu}=0.04\eqno(83)$$
With the decrease of $v_L$ the value $\tan\beta$ comes closer and closer to
1, while the increase of $v_L $ results in the removal of the values of
$\tan\beta$ from 1 to 0. The increase of the heavy neutrino masses
does not practically influence the behaviour of the contours shown in Fig.7.
At the increase of $f_{\mu\mu}$ there is the reduction of the allowed values
of $\tan\beta$ with the growth of the $m_{S_1}$ values. For example, when
$f_{\mu\mu}=0.1$ (all the remaining parameters are unchanged), then at
$m_{S_1}=165$ GeV the values of $\tan\beta$ lie within the interval
(0.0015, 0.4542) and at $m_{S_1}=200$ GeV those lie within the interval
(0.0211, 0.5562).

\section*{3. Conclusions}

\hspace*{8mm}
We have considered the Higgs sector of the LRM as a source of the muon €ŒŒ
value observed at BNL.
The contributions from the interactions of the doubly charged Higgs
bosons ($\Delta^{(--)}_{1,2}$), the singly charged ($h^{(-)}$ and
$\tilde{\delta}^{(-)}$) and the neutral ($S_1$) Higgs bosons both
with leptons and gauge bosons were taken into the account.
The found value of the muon €ŒŒ represents the function of
the Higgs boson masses and the Higgs boson couplings constants (CC's).
For the majority of the SM extensions the information about
the Higgs boson masses is at the level of knowledge of the lower borders
only. The situation with the CC's is even more pessimistic.
The experimental data derived up to now do not allow to obtain the
constraints on all the CC's. We managed to show that the most of part of
the CC's is the functions of the neutrino oscillation parameters.
By this it turned out that the values of these CC's are practically
not sensitive to the masses and the mixing angles in the light neutrino sector
and are mainly defined by the values of the heavy neutrino masses and by
the mixing angles between the light and the heavy neutrinos.
It should be particularly emphasized that this property
is common for all the models with the "see-saw" mechanism, i.e. for the models
with the heavy neutrino.

To explain the observed value of the muon AMM by the contributions either
from $S_1$ and $h^{(-)}$-bosons or from both of them, it is necessary to assume
that $\tan\beta$ is close to 1. To put this another way, the coincidence with
the BNL'00 result will take place at quasi-degeneracy of the bidoublet VEV's
($k_1\approx k_2 $), i.e. at the fine tuning of the bidoublet VEV 's.
As in this case obtained borders on the Higgs boson parameters weakly
depends on the neutrino sector parameters then the recovery of some information
concerning the masses and the mixing angles of the neutrinos will be rather
difficult. However the reverse side of this history is the fine capability for
detecting of $S_1$ and $h^{(-)}$-bosons. It is appeared that when
$\tan\beta$ is close to 1 then the values of the CC's for $h ^ {(-)} $- and
$S_1$-bosons are those that these bosons can be observed as the resonance
peaks in the whole series of the processes. For example, the $S_1$-boson
could be observed as the resonance splashes in the cross sections of the
reactions
$$\mu^+\mu^-\rightarrow\mu^+\mu^-,\tau^+\tau^-,\eqno(84)$$
$$\mu^+e^-\rightarrow\mu^-e^+,\eqno(85)$$
which practically have now background. The reactions (84) and (85) may be
investigated right now, because the energy of the muon beams used in the
current experiments is rather high. The Spin Muon Collaboration at CERN
has been working with muon beams having energy 190 GeV [47] and the FNAL
experiments investigating the muon-proton interaction has been using
the muons with energies of 470 GeV [48]. The reactions (84) and (85) can be
also investigated at the muon colliders (MC's) which are now under design.
For the detection of the $h^{(-)}$-bosons one could employ the reactions
$$e^-\nu_e\rightarrow W^-_1Z_1,\eqno(86)$$
$$e^-\nu_e\rightarrow \mu^-\nu_{\mu},\eqno(87)$$
which have the $s$-channel diagrams with the exchange of the $h^{(-)}$-boson
[39]. The ultra high energy cosmic neutrinos could be used for studying these
two reactions at such neutrino telescopes as BAIKAL NT-200, NESTOR and AMANDA.

However, one important point to remember that the fine tuning of the
parameters always belongs to the
extremely rare expedient which is used by Nature. For this reason
the variant with wider range of the parameters ensuring the agreement
between theory and experement is preferable.
The situation with the dominating contribution to the muon AMM from
$\Delta^{(--)}_{1,2}$- and $\tilde{\delta}^{(-)}$-bosons
is just such a case. However and in this case it is still far to the final
definition of the heavy neutrino parameters. Having established the values of
$f_{\mu\mu}$, $v_L $ and $v_R$ we shall obtain only two equations for the
definition of the quantities $\varphi_{\mu}$, $\theta_N$, $m_{N_{\mu}}$ and
$m_{N_{\tau}}$, that is obviously not enough.
Certainly it is possible to select the conventional way too.
Out of the five parameters of the heavy neutrino sector ($\varphi_{\mu}$,
$\varphi_{\tau}$, $\theta_N$, $m_{N_{\mu}}$, $m_{N_{\tau}}$)
we may fix the four parameters and vary the one, say, $m_{N}$.
At such an approach instead of the contours
shown in Fig.5 we shall have the contours constructed in $m_{N_{\mu}}$
{\it{vs.}} $m_{\Delta_2}$ parameter space,
that will not introduce anything essentially new to our analysis.
The most important here is something else, namely, when the contribution to the
muon €ŒŒ is truly caused by the $\Delta^{(-)}_1$-, $\tilde{\delta}^{(-)}$-
and/or $\Delta^{(-)}_2$-bosons then the further way of
defining the parameters of the heavy neutrino without their direct observation
is evident. For example, we could investigate the reactions
$$\mu^-\mu^-\rightarrow\mu^-\mu^-,\eqno(88)$$
$$\mu^-\mu^-\rightarrow\mu^-\tau^-,\eqno(89)$$
$$\mu^-\mu-\rightarrow\tau^-\tau^-,\eqno(90)$$
which may be observed at the MC's. All these reactions are going through
the $s$-channels with the exchanges of the $\Delta^{(--)}_{1,2}$-bosons.
Therefore, their cross sections have two resonance peaks related to the
Higgs bosons. Detecting of the reaction (88)
will allow to determine $f_{\mu\mu}$, while the invesigation of reactions
(89) and (90)
will yield the information about $f_{\mu\tau}$ and $f_{\tau\tau}$ respectively.
Then the use of Eqs. (51) --- (55) will allow to define the regions in which
the values of the heavy neutrino masses and the mixing angles are
constrained.

\section*{References}
\begin{description}
\item[1] R. M. Carey {\it{et al.}}, Muon g-2 Collaboration, Phys. Rev. Lett.
{\bf{82}}, 1632, 1999.
\item[2] H.N.Brown {\it{et al.}}, Muon g-2 Collaboration, Phys. Rev.
{\bf{D62}}, 091101, 2000.
\item[3] H.N.Brown {\it{et al.}}, Muon g-2 Collaboration, Phys. Rev. Lett.
{\bf{86}}, 2227, 2001.
\item[4] C.Caso {\it{et al.}}, Particle Data Group, Eur.Phys.J. {\bf{C3}}, 1,
1998.
\item[5] P. Mohr and B.Taylor, Rev. Mod. Phys. {\bf{72}} 351 (2000).
\item[6] A. Czarnecki and W.J.Marciano, Phys. Rev. {\bf{D64}} 013014 (2001).
\item[7] M. Davier and A. Hocker, Phys. Lett. {\bf{B435}}  427 (1998);
M.Davier, Nucl.Phys. (Proc.Suppl.) {\bf{B76}} 327 (1999).
\item[8] B.Krause, Phys. Lett. {\bf{B390}} 392 (1997);
R.Alemany, M. Davier, and A. Hocker, Eur. Phys. J. {\bf{C2}}  123 (1998).
\item[9] H. Hayakawa, T. Kinoshita and A. Sanda, Phys. Rev. Lett. {\bf{75}}
790 (1995); M. Hayakawa and T. Kinoshita, Phys. Rev. {\bf{D57}} 465(1998).
\item[10] J. Bijnens, E. Pallante and J. Prades, Phys.
Rev. Lett. {\bf{75}} 3781 (1995); E. Nucl. Phys. {\bf{B474}} 379 (1996).
\item[11] M. Knecht, A. Nyffeler, M. Perrottet and E. de Rafael, Phys. Rev.
Lett. {\bf{88}} 071802 (2002).
\item[12] M. Hayakawa and T. Kinoshita, arXiv:hep-ph/0112102.
\item[13] J. Bijnens, E. Pallante and J. Prades, arXiv:hep-ph/0112255.
\item[14] H.N.Brown {\it{et al.}}, Muon g-2 Collaboration, hep-ex/0208001
(is accepted for publicaton in Phys. Rev.).
\item[15] F.Jegerlehner, talk at Conference on "Hadronic Conributions to the
Anomalous Magnetic Moment of the Muon", Marseille, March 2002.
\item[16] K.Hagiwara, A.D.Martin, D.Nomura and T.Teubner,
T.Teubner talk at SUSY02, DESY Hamburg, June 2002.
\item[17] R.R. Akhmetshin {\it{et al.}} CMD2 Collaboration,
Phys. Lett. {\bf{B527}} 161 (2002);
J.Z.Bai {\it{et al.}}, BES Collaboration, Phys. Rev. Lett. {\bf{84}} 594 (2000);
J.Z.Bai {\it{et al.}}, BES Collaboration, Phys. Rev. Lett. {\bf{88}} 101802
(2002); M.N.Achanov {\it{et al.}}, SND Collaboration, arXiv:hep-ex/9809013.
\item[18] A. Nyffeler,arXiv:hep-ph/0203243.
\item[19] S.Fukuda {\it{et al.}},  Super-Kamiokande Collaboration,
Phys. Rev. Lett. {\bf{86}}, 5656 (2001);
Q.R.Ahmad {\it{et al.}},  SNO Collaboration, Phys. Rev. Lett. {\bf{87}},
071301, 2001.
\item[20] G.P.Zeller {\it{et al.}}, NuTeV Collaboration, hep-ph/0110059;
K.S.McFarland {\it{et al.}}, NuTeV Collaboration, hep-ex/0210010.
\item[21] H.V.Kladpor-Kleingrothaus and U.Sarkar, Mod. Phys. Lett. {\bf{A1}}6
2469 (2001).
\item[22] A. Czarnecki, W.J.Marciano// Alberta Thy 03-01, BNL-HET-01/4,
hep-ph/0102122.
\item[23] P.Das {\it{et al.}}, hep-ph/0102242; S.N.Gninenko and N.V.Krasnikov,
hep-ph/0102222.
\item[24] M.Beccaria {\it{et al.}}, Phys. Lett. {\bf{B448}}, 129, 1999.
\item[25] D.Chakraverty {\it{et al.}}, hep-ph/0102180; U.Mahanta,
hep-ph/0102176.
\item[26] R.Casadio {\it{et al.}}, Phys. Lett. {\bf{B495}}, 378, 2000;
M.L.Graesser, Phys. Rev. {\bf{D61}}, 074019, 2000.
\item[27] M.C.Gonzales-Garcia and S.F.Novaes, Phys. Lett. {\bf{B389}}, 707, 1996;
K.Lane, hep-ph/0102131.
\item[28] R.T.Huang {\it{et al.}}, IHEP Beijing preprint 2001, hep-ph/0102193;
R.T.Huang {\it{et al.}}, hep-ph/0103193.
\item[29] D.Choudhury {\it{et al.}}, HRI preprint 2001, hep-ph/0102199.
\item[30] K.Lane, BUHEP-01-3, 2001, hep-ph/0102131.
\item[31] D.Colladay and V.A.Kostelecky, Phys.Rev. {\bf{D58}}, 116002, 1998.
\item[32] W.Liu {\it{et al.}}, Phys. Rev. Lett. {\bf{82}}, 711, 1999.
\item[33] D.E.Groom {\it{et al.}}, Particle Data Group, Eur. Phys. J.
{\bf{C15}}, 1, 2000.
\item[34] J.P.Leveille, Nucl. Phys. {\bf{B137}}, 63, 1978.
\item[35] M.Sher, Physics Reports, {\bf{179}} (1989) 273.
\item[36] H.P. Nilles, Phys. Rep. {\bf{110}} (1984) 1; H. Haber and G.Kane,
Phys. Rep. {\bf{117}} (1985) 75.
\item[37] Singer M, Valle J.W.F. and Scechter J., Phys. Rev.
{\bf{D22}} (1980) 738; Singer M, Valle J.W.F. and Singer M, Phys. Rev.
{\bf{D28}} (1983) 540.
\item[38] N.G.Deshpande {\it{et al.}}, Phys. Rev. {\bf{D44}}, 837, 1989.
\item[39] G.G.Boyarkina, O.M.Boyarkin, Eur.Phys. J. {\bf{C13}}, 99, 2000.
\item[40] G.G.Boyarkina, O.M.Boyarkin, Yad. Fiz. {\bf{61}}, 757, 1998.
\item[41] O.M.Boyarkin and T.I.Bakanova, Phys. Rev. {\bf{D62}}, 075008, 2000.
\item[42] G.G.Boyarkina, O.M.Boyarkin, Yad. Fiz. {\bf{60}}, 683, 1997.
\item[43] T.G.Rizzo, Phys. Rev., {\bf{D25}} (1982) 1355.
\item[44] G.Abbiendi, {\t{et al.}}, OPAL Collaboraion, CERN-EP-2001-028,
hep-ph/0111059.
\item[45] R.Barate {\it{et al.}}, ALEPH Collaboraion, Phys. Lett. {\bf{B495}},
1, 2000;
P.Abreu {\it{et al.}}, DELPHI Collaboraion, Phys. Lett. {\bf{B499}}, 23, 2001;
M.Acciarti {\it{et al.}}, L3 Collaboraion, hep-ex/0012019;
G.Abbiendi {\it{et al.}}, OPAL Collaboraion,Phys. Lett. {\bf{B499}}, 38, 2001.
\item[46] A.Dedes, H.E.Haber//hep-ph/0102297.
\item[47] SMC, D.Adams et al., Phys. Lett., {\bf{B357}} (1995) 248.
\item[48] M.R.Adams et al., The E665 Collaboration, FERMILAB-PUB-97/103-E94.
E665, 1997.
\end{description}

\newpage
\section*{Fugures Captions}
{\bf{Fig.1}}. One-loop diagrams contribute to the muon AMM due to
the doubly charged Higgs bosons $\Delta^{(--)}_{1,2}$. The wavy
line represent
the electromagnetic field.\\
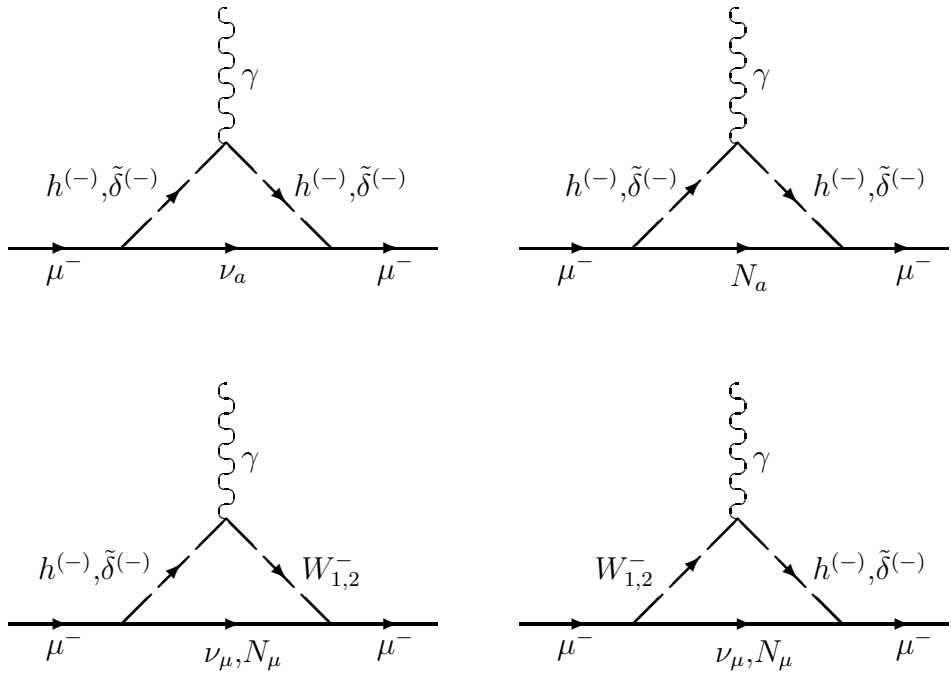
{\bf{Fig.2}}. One-loop diagrams contribute to the muon AMM due to
the singly charged Higgs bosons $\tilde{\delta}^{(-)}$ and $h^{(-)}$.\\
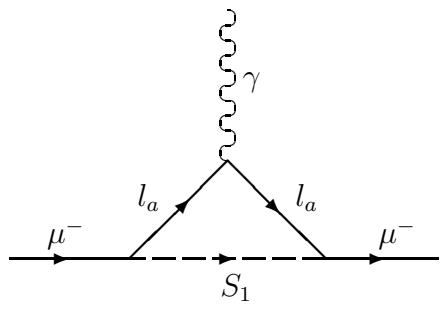
{\bf{Fig.3}}. One-loop diagrams contribute to the muon AMM due to
the lightest neutral Higgs boson $S_1$.\\
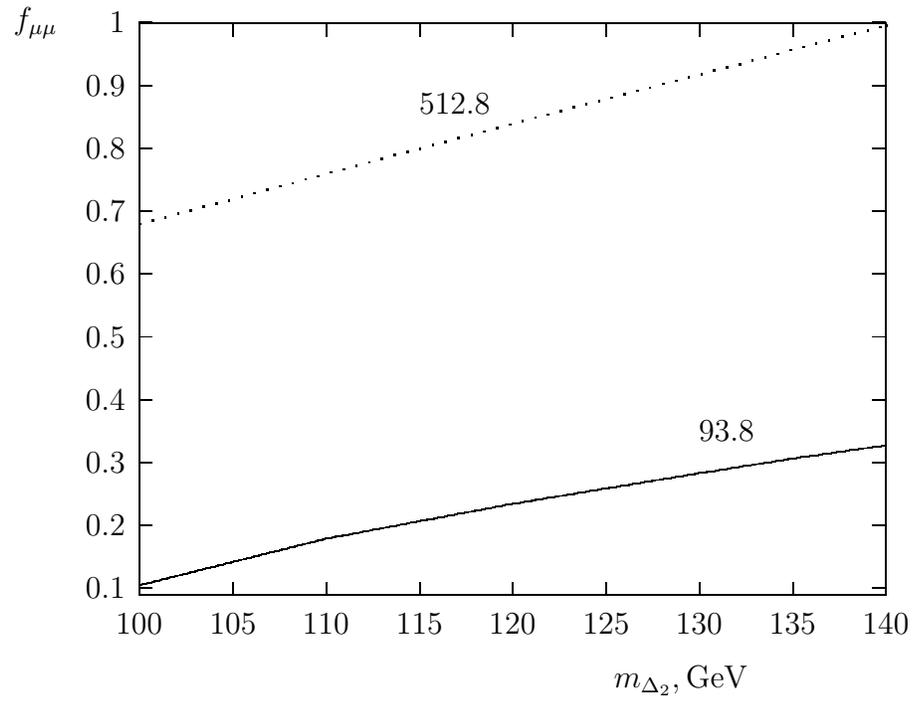
{\bf{Fig.4}}. Contours of the one-loop contribution from the
$\Delta^{(--)}_2$-boson to the muon AMM. \\
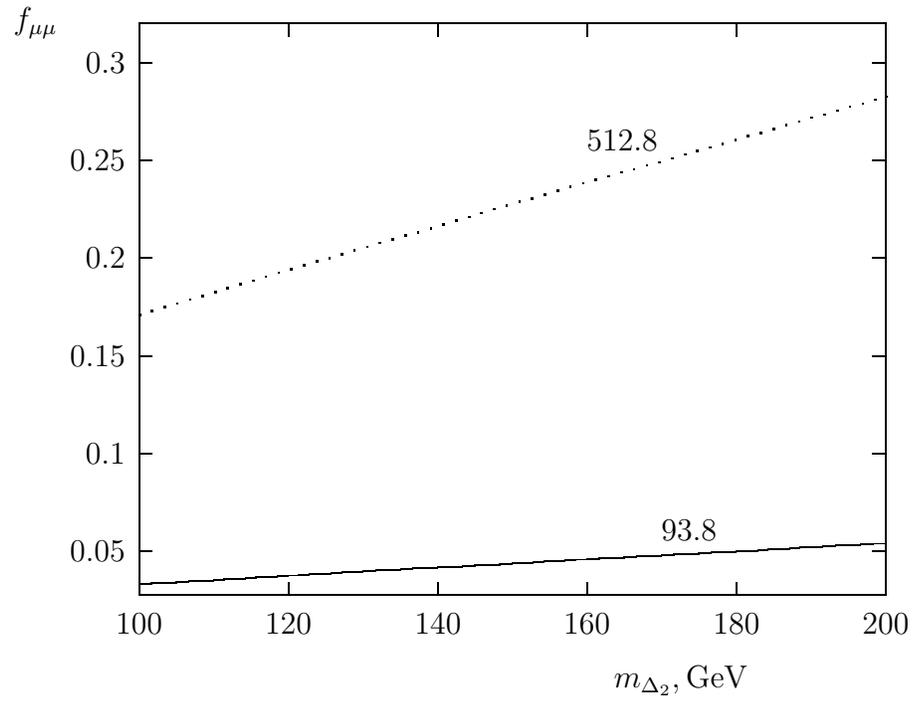
{\bf{Fig.5}}. Contours of the one-loop contribution from the
$\Delta^{(--)}_{1,2}$ and the $\tilde{\delta}^{(-)}$-boson to the muon AMM.\\
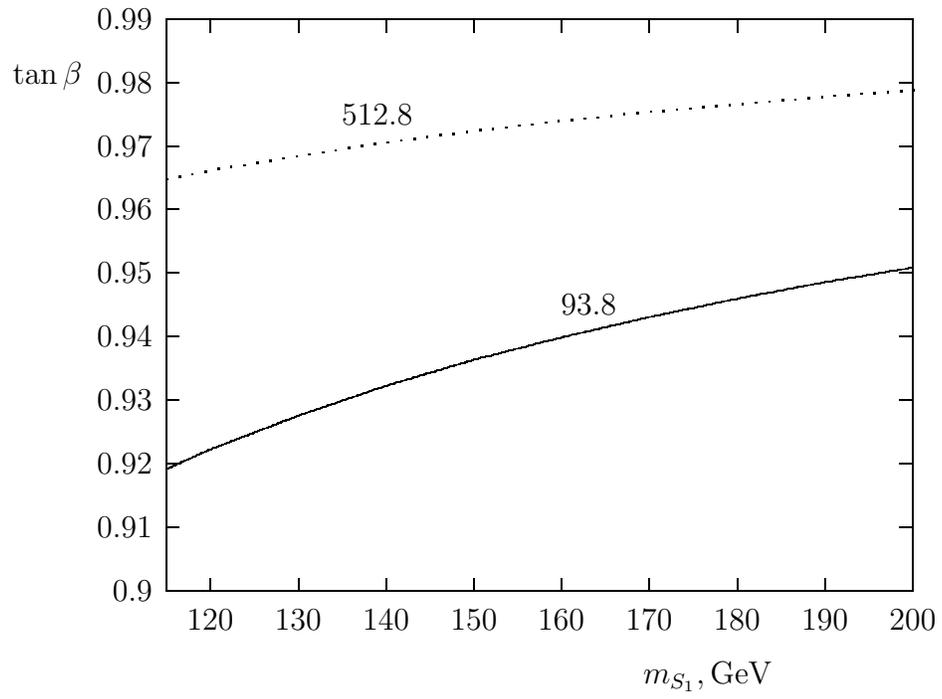
{\bf{Fig.6}}. Contours of the one-loop contribution from the
$S_1$-boson to the
muon AMM.\\
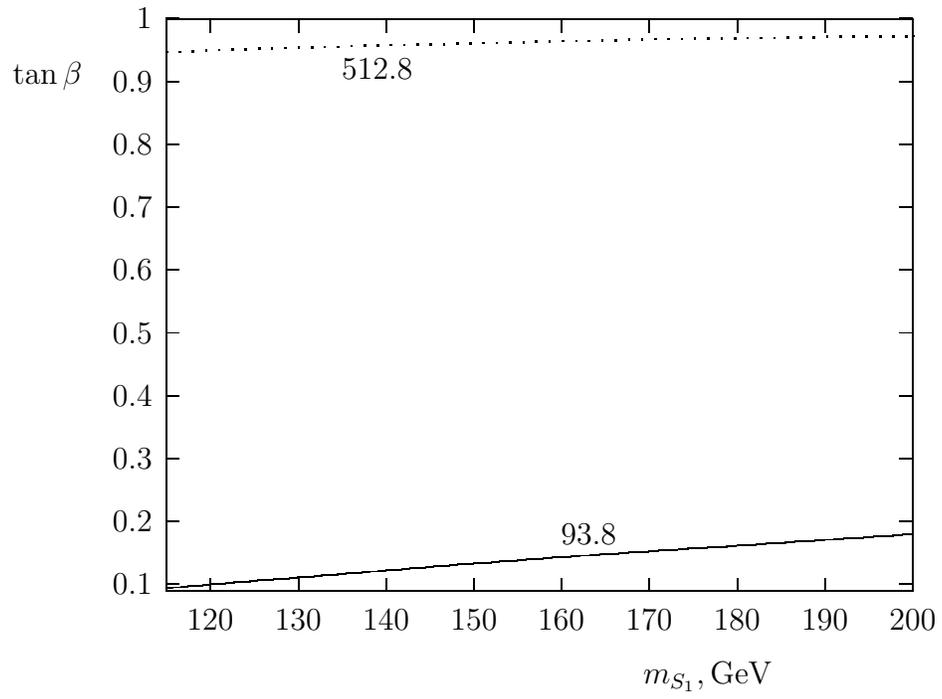
{\bf{Fig.7}}. Contours of the one-loop contribution from the
$S_1$- and
$h^{(-)}$-boson to the muon AMM.\\

%\newpage
\begin{figure}
\setlength{\unitlength}{1 mm}

\hspace*{7cm}{Fig. 7}

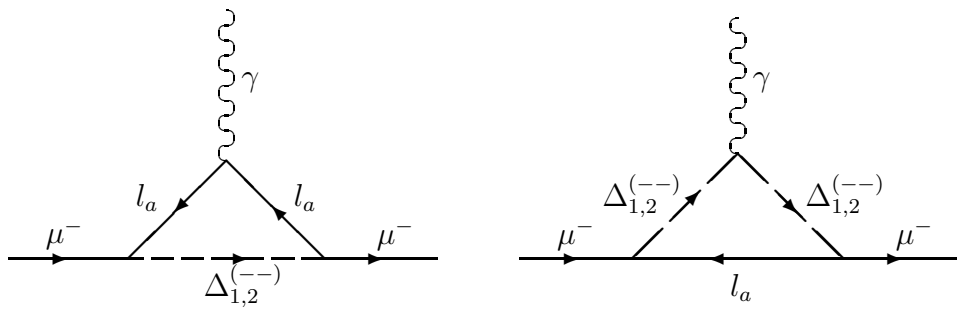
\end{document}